\documentclass[runningheads]{llncs}
\usepackage[T1]{fontenc}
\usepackage{graphicx}
\usepackage{hyperref}

\hypersetup{
    colorlinks=true,
    linkcolor=black,
    citecolor=black,
    urlcolor=blue
}

\begin{document}
\title{Agents of ViTAL: Ethics Missions --- A Narrative-Centered Learning Environment with a Co-Designed Conversational Agent for Middle School AI Ethics}
\titlerunning{Agents of ViTAL: Ethics Missions}
%
%
%

\author{Sarah Burriss\inst{1} \and
Jinzhi Zhou\inst{2} \and
Namrata Srivastava\inst{1} \and
Justin Phillips\inst{3} \and 
Courtney Barron\inst{3} \and
Kara Cassell\inst{3} \and
Albert Na\inst{1} \and
Megan Humburg\inst{2} \and
Tom Hu\inst{1} \and
Benjamin Yang\inst{1} \and
Corey Brady\inst{4} \and
Ole Molvig\inst{1}
}
\authorrunning{S. Burriss et al.}
%
\institute{Vanderbilt University, USA \and
Indiana University, USA \and 
North Carolina State University, USA  \and
Southern Methodist University\\
\email{sarah.burriss@vanderbilt.edu}\\
}

\maketitle              
\begin{abstract}
\textbf{Agents of ViTAL: Ethics Missions} is a browser-based, narrative-centered learning environment in which middle school students collaboratively evaluate whether a fictional school should adopt an AI-powered classroom feedback tool. Students investigate stakeholder perspectives, weigh tradeoffs across three AI ethics dimensions (privacy, bias, and environmental impact), and negotiate group consensus through a shared Ranking Challenge interface. The environment embeds EthicsBot, a conversational agent designed as a peer-like thought partner that scaffolds ethical reasoning and collaborative discussion. The project is being iteratively co-designed with high school students, whose feedback shapes EthicsBot's role, behavior, and guardrails. Initial classroom implementations with four ninth-grade classes demonstrated strong engagement and substantive ethical reasoning grounded in students' lived experiences with AI. The demo invites attendees to explore the learning environment, participate in the collaborative Ranking Challenge, and interact with EthicsBot.


\keywords{AI ethics \and narrative-centered learning \and conversational agents \and co-design \and collaborative learning \and K–12 \and LLM}
\end{abstract}
\noindent\textbf{Demo Video:} \url{{https://vimeo.com/1195380607}} \\

\noindent\textbf{System Overview: }

\noindent Agents of ViTAL: Ethics Missions is a narrative-centered learning environment~\cite{lester2012narrative} where students collaboratively investigate the ethical risks and benefits of adopting an AI-powered classroom tool in a fictional school. The system combines stakeholder-based inquiry, collaborative decision-making, and an embedded conversational agent, EthicsBot, that supports group ethical reasoning.

\noindent\textbf{Key Components:} \\
\noindent\textbf{\textit{Agent Orientation in the Libratory:}}
Students begin by entering the “Libratory,” a virtual hub space combining elements of a library and laboratory. After selecting avatars, students meet HelperBot and Professor Mugo, who introduce the mission: evaluating whether Simulation Middle School should adopt PresenterPal, an AI-powered classroom feedback system.


\setlength{\intextsep}{5pt}
\begin{figure}
    \centering
    \includegraphics[width=0.82\linewidth]{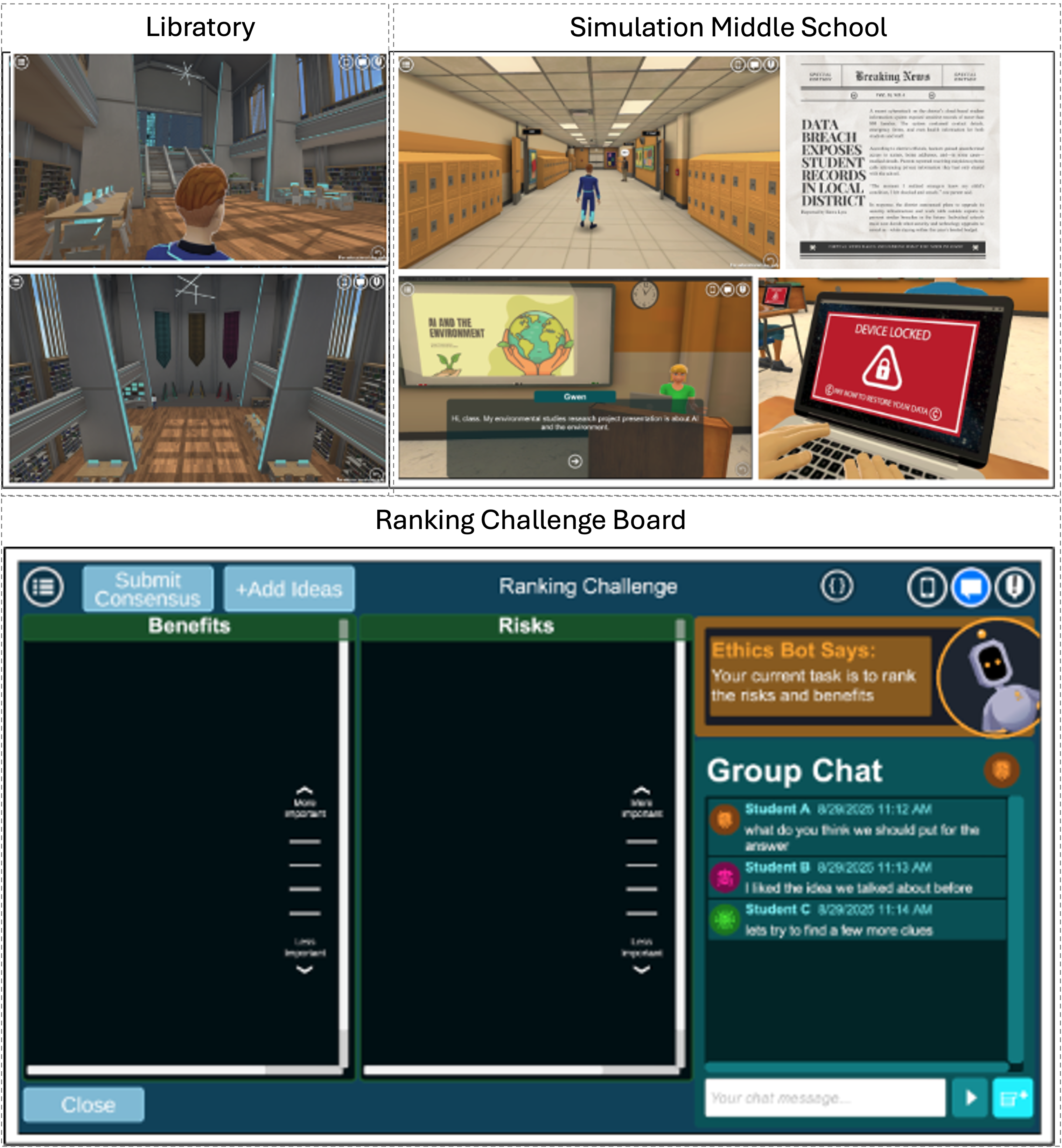}
    \caption{Key components of the Agents of ViTAL learning environment}
    \label{fig:placeholder}
\end{figure}

\noindent\textbf{\textit{Investigation at Simulation Middle School:}}
Students then explore Simulation Middle School to gather perspectives from stakeholders including teachers, students, administrators, and IT staff. They collect evidence artifacts (e.g., newspaper clippings about data breaches) and encounter emergent scenarios that surface tradeoffs across three ethics dimensions: privacy, bias, and environmental impact. 
For example, when configuring PresenterPal, students must decide whether student interaction data should be stored locally or on cloud servers, a decision linked to both the school’s limited budget and cybersecurity risks.

\noindent\textbf{\textit{Ranking Challenge Board:}}
After the investigation phase, students collaboratively complete the Ranking Challenge. In this interface, groups identify and rank AI risks and benefits according to importance while negotiating their reasoning through an integrated group chat. The shared ranking board functions as a boundary object that visually represents the group’s evolving ethical stance and supports collaborative consensus-building. The activity encourages students to articulate competing viewpoints, justify tradeoffs using evidence gathered during gameplay, and collectively produce a final recommendation regarding PresenterPal adoption.


\noindent\textbf{\textit{EthicsBot:}}
EthicsBot provides scaffolded conversational support during collaborative discussion. Rather than supplying answers, the agent prompts students to consider alternative perspectives, clarify reasoning, and reflect on ethical tradeoffs. The current version uses scripted scaffolds, while an LLM-backed version is being iteratively developed through student co-design. Student feedback has shaped the agent’s role as a peer-like thought partner, including its scaffolding strategies and guardrails around over-reliance and hallucination. 

\vspace{5pt}
\noindent\textbf{Originality and Strengths of the System:}

\noindent Three features distinguish Agents of ViTAL within the AIED landscape: 

\noindent(1) While AI ethics curricula for K–12 learners are emerging, few systems embed ethical reasoning within a narrative-centered collaborative environment featuring stakeholder simulation, dynamic consequences, and authentic tradeoffs grounded in school contexts ~\cite{chang2025systematic,ma2025fostering,wiese2025ai}. Students are positioned as active participants in an AI adoption decision rather than passive recipients of ethics instruction. 

\noindent(2) Students serve as co-designers of EthicsBot itself. Their feedback directly shapes agent's conversational behavior, scaffolding strategies, and safety guardrails, centering learner voice in the development of an educational AI system.

\noindent(3) The environment operationalizes the ``From Tools to Teammates'' conference theme through EthicsBot's design: the agent sustains student reasoning, supports constructive disagreement, and encourages collaborative negotiation. 

A classroom pilot with 80 consented ninth-grade students across four classes yielded strong engagement and substantive ethical reasoning, with students connecting gameplay scenarios to their own experiences with AI technologies.


\begin{credits}
\subsubsection{\ackname}
This work was supported by the National Science Foundation under award DRL-2112635 and a LIVE Ignite Tech Grant from Vanderbilt University.
 \end{credits}

%
%
%
\bibliographystyle{splncs04}
\bibliography{references}
%


\end{document}